\documentstyle[psfig]{caosp201}

\begin{document}
\pubyear{2002}
\volume{32}
\firstpage{1}
\htitle{Photometry of symbiotic stars X.}
\hauthor{A. Skopal et al.}
\title{Photometry of symbiotic stars}
\subtitle{X. EG\,And, Z\,And, BF\,Cyg, CH\,Cyg, V1329\,Cyg, \\
             AG\,Dra, RW\,Hya, AX\,Per and IV\,Vir}
\author{A.\,Skopal\inst{1} 
\and 
  M.\,Va\v nko\inst{1} 
\and 
  T.\,Pribulla\inst{1}
\and
  M.\,Wolf\inst{2\,\star}\footnotetext{$^{\parbox{0.5cm}{$\star$}}$Visiting 
                Astronomer, San Pedro Observatory}
\and
  E.\,Semkov\inst{3}
\and 
  A.\,Jones \inst{4}} 
\institute{
  \lomnica
\and
  Astronomical Institute, Charles University Prague, CZ-180 00 Praha 8, 
  \mbox{V Hole\v{s}ovi\v{c}k\'ach} 2, Czech Republic 
\and
  Institute of Astronomy, Bulgarian Academy of Sciences, Tsarigradsko 
  shose Blvd. 72, Sofia 1784, Bulgaria
\and
  Carter Observatory, PO Box 2909, Wellington 1, New Zealand}
\date{December 15, 2001}

\maketitle

\begin{abstract}
We present new photometric observations of EG\,And, Z\,And, BF\,Cyg, 
CH\,Cyg, V1329\,Cyg, AG\,Dra, RW\,Hya, AX\,Per and IV\,Vir made in 
standard Johnson $UBVR$ system. The current issue summarizes observations 
of these objects to 2001 December. The main results can be summarized 
as follows:
{\bf EG\,And}: 
A periodic double-wave variation in all bands as a function 
of the orbital phase was confirmed. A maximum of the light changes 
was observed in $U$ ($\Delta U \sim 0.5$\,mag). 
{\bf Z\,And}: 
Our observations cover an active phase, which peaked around 8.4 in $U$ 
at the beginning of 2000 December. Consequently, a gradual decrease in 
the star's brightness has been observed. 
{\bf BF\,Cyg}: 
Periodic wave-like variation in the optical continuum reflects 
a quiescent phase of this star. A complex light curve (LC) profile 
was observed. 
{\bf CH\,Cyg}:
The recent episode of activity ended in Spring of 2000. We determined 
position of an eclipse in the outer binary at JD~2\,451\,426$\pm 3$. 
Recent observations indicate a slow increase in the star's brightness. 
{\bf V1329\,Cyg}:
Observations were made around a maximum at 2001.2. 
{\bf AG\,Dra}:
Our measurements from the Autumn of 2001 revealed a new eruption, which
peaked at $\sim$JD~2\,452\,217. 
{\bf RW\,Hya}:
The light minimum in our mean visual LC precedes the time of 
the spectroscopic conjunction of the giant in the binary. 
{\bf AX\,Per}:
Periodic wave-like variation was observed. Our recent observations 
revealed a secondary minimum at the orbital phase 0.5, seen best in 
the $V$ and $B$ bands. 
{\bf IV\,Vir}:
The LC displays a double-wave throughout the orbital cycle. 
%The 
%profile is similar to that measured in the Str\"omgren $y$ band. 
%
\end{abstract}

\section{Introduction}

Symbiotic stars are long-period (typically 1-3\,years) interacting 
binaries consisting of a red giant and a hot compact object. 
The giant component loses mass, part of which is accreted by
its companion. During {\em quiescent phases} the hot star ionizes
a portion of the giant's wind, giving rise to nebular emission.
During {\em active phases} the hot star expands in radius and becomes
significantly cooler. An increase in the star's brightness, 
typically by 2-3\,mag, represents the most impressive face of such 
active stages.

The primary aim of our monitoring programme is to present LCs of 
well-studied symbiotic stars, and thus to provide a good basis 
for their detailed future studies. 
In this respect, this paper continues 
the work of Skopal et al. (2000a), Skopal (1998) and other 7 campaign's 
papers on the photometry of symbiotic stars, originally launched 
by Hric \& Skopal (1989). 

\section{Observations}

The majority of the $U,~B,~V,~R$ measurements were performed 
in standard Johnson system using single-channel photoelectric 
photometers mounted in the Cassegrain foci of 0.6-m reflectors at 
the Skalnat\'{e} Pleso (hereafter SP in Tables) and Star\'{a} Lesn\'{a} 
observatories (SL). Larger uncertainties (about 0.1\,mag) during some 
nights are marked in Tables by '::'. Further details about observation 
procedure are given in Skopal, Hric \& Urban (1990) and 
Hric et al. (1991). 

$U,~B,~V$ observations of RW\,Hya and IV\,Vir were secured at 
the San Pedro M\'artir Observatory, Baja California, Mexico (M), during 
two weeks in June 2001. The 0.84-m Cassegrain reflector equipped with 
the photon-counting photometer Cuentapulsos (utilizing a RCA 31034
photomultiplier) was used. These observations consisted from 10-second 
integrations in each filter. They were carefully reduced to the standard 
$UBV$ system and corrected for differential extinction using the reduction 
program {\scriptsize HEC}~22 rel.13.2 (Harmanec \& Horn 1998).

Some observations in the $B$ and $V$ band were made with the 50/70/172\,cm 
Schmidt telescope of the National Astronomical Observatory Rozhen, 
Bulgaria (R). A CCD camera SBIG ST-8 and Johnson-Cousins set of filters 
were used. The chip of the camera is KAF 1600 (16 bit), with dimensions 
of 13.8$\times$9.2\,mm or 1530$\times$1020 pixels. The size of
the pixel is 9$\times$9\,$\mu$m and the scale of 1".1/pixel. 
The readout noise was 10 ADU/pixel and the gain 2.3 e-/ADU. 
All frames were dark subtracted and flat fielded. Photometry was 
made with {\scriptsize DAOPHOT} routines.

In addition, 783 and 776 visual magnitude estimates of RW\,Hya and IV\,Vir, 
respectively, were obtained during 1989.0 -- 2001.7 by one of us (AJ) 
with a private 12".5 f/5 reflector. 

\section{Results}

\subsection{EG\,And}

We measured EG\,And (HD\,4174, BD+39\,167) with respect 
to HD\,4143 (SAO\,63\-173, BD+37\,2318). 
To obtain magnitudes in $B$ and $V$ we used the standard star 
HD\,3914 ($V$ = 7.00, $B-V$ = 0.44) and conversion between both 
stars, HD4143 -- HD3914 = 4.640, 2.722 and 1.563 in the $U$, 
$B$ and $V$ band, respectively (Hric et al. 1991). 
For the CCD measurements, stars SAO\,36021 and HD\,4127 were also 
used as comparison and check star. 
The data are compiled in Table 1. Figure 1 shows the phase diagram 
for magnitudes in the $U$ band from Table 1. A modulation displaying 
a double wave during the orbital period is the most significant 
feature of the $U$-LC. A large amplitude, $\Delta U_{max} \sim $0.5\,mag, 
is difficult to explain by the ellipsoidal variation due to tidal 
distortion of the red giant component (Wilson \& Vaccaro 1997).
The nature of such type of variability is probably given by the shape 
of the nebula in the binary. A small and partially optically thick 
nebula can cause a double-wave variation in LCs as a function of 
the orbital phase and thus mimic the ellipsoidal variation 
(see Skopal 2001 for more detail). 
%
%================================|
%           Fig. 1: EG And       |
%================================|
%\vspace*{0.25cm}
\begin{figure}[!ht]
  \centering
  \centerline{\hbox{
  \psfig{figure=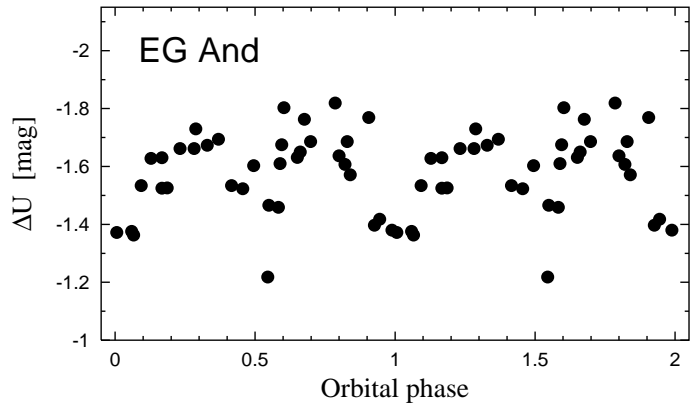,width=9cm,angle=-90}}}
%\vspace{0.25cm}
\caption[ ]{The LC of EG\,And in $U$. It consists of the data published 
in this paper only (Table 1). The data were folded with the ephemeris for 
the primary minima, $Min = JD\,2\,446\,336.7 + 482\times E$ (Skopal 1997).
}
\end{figure}
%
%
%==============================|
%       Table 1: EG And        |
%==============================|
\begin{table}[!ht]
%\small
\scriptsize
\begin{center}
\caption{$U,~B,~V,~R$ observations of EG\,And}
\begin{tabular}{lccccccc}
\hline
Date & JD~24... & Phase$^{\star}$ & $\Delta U$ & $B$ & $V$ & $\Delta R$ & Obs \\
\hline
 Sep 03, 98 & 51060.429 & 0.800 & -1.637 &  8.884 &  7.239 & -1.446 &SP  \\
 Sep 22, 98 & 51079.419 & 0.840 & -1.571 &  8.851 &  7.208 & -1.466 &SP  \\
 Nov 12, 98 & 51130.380 & 0.945 & -1.418 &  8.874 &  7.208 & -1.465 &SP  \\
 Dec 03, 98 & 51151.385 & 0.989 & -1.380 &  8.884 &  7.230 & -1.441 &SP  \\
 Dec 11, 98 & 51159.434 & 0.006 & -1.372 &  8.867 &  7.197 & -1.474 &SP  \\
 Jan 06, 99 & 51185.227 & 0.059 & -1.376 &  8.877 &  7.226 & -1.442 &SP  \\
 Jan 22, 99 & 51201.351 & 0.093 & -1.534 &  8.816 &  7.177 & -1.481 &SP  \\
 Feb 27, 99 & 51237.290 & 0.167 & -1.630 &  8.844 &  7.200 & --     & SL \\
 Sep 16, 99 & 51437.540 & 0.583 & -1.459 &  8.838 &  7.182 & -1.496 &SP  \\
 Oct 30, 99 & 51482.371 & 0.676 & -1.763 &  8.729 &  7.080 & --     & SL \\
 Dec 22, 99 & 51535.315 & 0.786 & -1.819 &  8.732 &  7.094 & -1.567 &SP  \\
 Jan 08, 00 & 51552.258 & 0.821 & -1.607 &  8.826 &  7.171 & -1.498 &SP  \\
 Jan 12, 00 & 51556.220 & 0.829 & -1.686 &  8.798 &  7.128 & --     & SL \\
 Feb 18, 00 & 51593.231 & 0.906 & -1.769 &  8.840 &  7.181 & --     & SL \\
 Feb 28, 00 & 51603.245 & 0.926 & -1.397 &  8.827 &  7.148 & --     & SL \\
 Aug 17, 00 & 51774.481 & 0.282 & -1.662 &  8.819 &  7.143 & --     & SL \\
 Aug 20, 00 & 51777.480 & 0.288 & -1.730 &  8.773 &  7.123 & -1.545 &SP  \\
 Sep 09, 00 & 51797.328 & 0.329 & -1.673 &  8.812 &  7.151 & --     & SL \\
 Sep 29, 00 & 51816.561 & 0.369 & -1.694 &  8.787 &  7.155 & -1.499 &SP  \\
 Oct 21, 00 & 51839.334 & 0.416 & -1.534 &  8.833 &  7.155 & --     & SL \\
 Oct 29, 00 & 51847.441 & 0.433 &   --   &  8.81  &  7.34  & --     & R  \\
 Oct 20, 00 & 51848.451 & 0.435 &   --   &  8.87  &   --   & --     & R  \\
 Nov 09, 00 & 51858.417 & 0.456 & -1.523 &  8.845 &  7.199 & -1.483 &SP  \\
 Nov 28, 00 & 51877.464 & 0.495 & -1.603 &  8.876 &  7.233 & -1.457 &SP  \\
 Dec 22, 00 & 51901.270 & 0.545 & -1.218 &  8.937 &  7.228 & --     & SL \\
 Dec 24, 00 & 51903.214 & 0.549 & -1.466 &  8.927 &  7.229 & -1.439 &SP  \\
 Dec 24, 00 & 51903.358 & 0.549 &   --   &  8.86  &  7.42  &  --    & R  \\
 Jan 12, 01 & 51922.403 & 0.589 & -1.610 &  8.834 &  7.185 & -1.490 &SP  \\
 Jan 15, 01 & 51925.359 & 0.595 & -1.675 &  8.816 &  7.170 & -1.510 &SP  \\
 Jan 19, 01 & 51929.262 & 0.603 & -1.803 &  8.813 &  7.138 & --     & SL \\
 Feb 11, 01 & 51952.272 & 0.651 & -1.631 &  8.820 &  7.143 & --     & SL \\
 Feb 16, 01 & 51957.230 & 0.661 & -1.650 &  8.793 &  7.146 & -1.543 &SP  \\
 Feb 17, 01 & 51958.302 & 0.663 &   --   &  8.72  &  7.23  &   --   & R  \\
 Mar 06, 01 & 51975.281 & 0.698 & -1.686 &  8.649 &  7.023 & -1.618 &SP  \\
 Aug 31, 01 & 52152.511 & 0.066 & -1.363 &  8.944 &  7.302 & -1.380 &SP  \\
 Sep 30, 01 & 52182.582 & 0.128 & -1.628 &  8.804 &  7.229 & -1.481 &SP  \\
 Oct 18, 01 & 52201.388 & 0.167 & -1.525 &  8.887 &  7.238 & -1.431 &SP  \\
 Oct 27, 01 & 52210.380 & 0.186 & -1.526 &  8.898 &  7.221 & -1.431 &SP  \\
 Nov 18, 01 & 52232.417 & 0.232 & -1.662 &  8.818 &  7.191 & -1.480 &SP  \\
\hline
\end{tabular}
\end{center}
$^{\star}$ $Min = JD\,2\,446\,336.7 + 482\times E$ (Skopal 1997)
\normalsize
\end{table}

\subsection{Z\,And}

The results of our photometric measurements of Z\,And (HD\,221650, 
BD+484093) are in Table 2. Stars SAO\,53150 
(BD+47\,4192; $V$ = 8.99, $B-V$ = 0.41, $U-B$ = 0.14) and
SAO\,63189 (BD+47\,4188; $V$ = 9.17, $B-V$ = 1.36, $U-B$ = 1.11), 
were used as a comparison and a check star, respectively. 
We obtained the magnitudes of both stars by their long-term 
measuring (1997 -- 1999) with respect to the previous comparison 
star (SAO\,35642; $V$ = 5.30, $B-V$ = -0.06, $U-B$ = -0.15). 
Resulting magnitudes of Z\,And were obtained only by using 
the first comparison (SAO\,53150) as the second one can be 
variable. CCD measurements were also compared to the star 
SAO\,53150 and, in addition, to SAO\,53133. 
A dominant feature of the presented LCs (Fig. 2) is a strong active 
phase, which began on 2000 August 31 (Skopal et al. 2000b). The star's 
brightness peaked around 8.4 in $U$ at the beginning of 2000 December 
(see also Sokoloski et al. 2001). Consequently, the brightness 
has been gradually decreasing. Comparing its latest magnitudes, 
$U\sim 9.7$, $B\sim 10.8$, $V\sim 10.0$, with those observed prior 
to the outburst ($U\sim 11.5$, $B\sim 12$, $V\sim 10.5$), indicates 
that Z\,And still remains at a high level of its activity. 
%
%==============================|
%         Fig. 2: Z And        |
%==============================|
%\vspace*{0.25cm}
\begin{figure}[!ht]
  \centering
  \centerline{\hbox{
  \psfig{figure=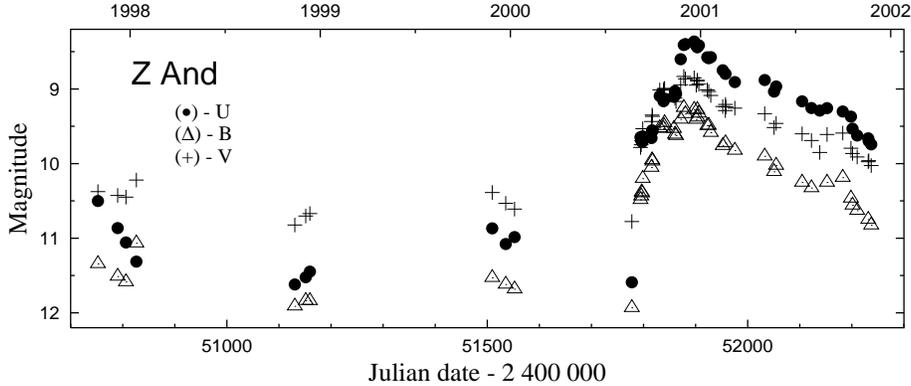,width=12cm,angle=-90}}}
%\vspace{0.25cm}
\caption[ ]{The $U,~B,~V$ LCs of Z\,And.}
\end{figure}
%
%
%==============================|
%       Table 2: Z And         |
%==============================|
\begin{table}[!ht]
%\small
\scriptsize
\begin{center}
\caption{$U,B,V,R$ observations of Z\,And}
\begin{tabular}{lccccccc}
\hline
Date & JD~24... & Phase$^{\star}$ & $U$ & $B$ & $V$ & 
                                            $\Delta R$  & Obs  \\
\hline
 Oct 30, 97 & 50752.390 & 0.654 & 10.501 & 11.344 & 10.375 &
3.671$^{\dagger}$ &SP  \\
 Dec 07, 97 & 50790.366 & 0.704 & 10.864 & 11.513 & 10.428 &
3.766$^{\dagger}$ &SP  \\
 Dec 23, 97 & 50806.321 & 0.726 & 11.057 & 11.587 & 10.449 &
3.826$^{\dagger}$ &SP  \\
 Jan 12, 98 & 50826.257 & 0.752 & 11.313 & 11.068 & 10.221 &  0.896 &SP  \\
 Nov 13, 98 & 51130.532 & 0.153 & 11.620 & 11.912 & 10.823 &
4.167$^{\dagger}$ &SP  \\
 Dec 03, 98 & 51151.325 & 0.180 & 11.524 & 11.838 & 10.702 &
4.090$^{\dagger}$ &SP  \\
 Dec 11, 98 & 51159.378 & 0.191 & 11.447 & 11.841 & 10.670 &
4.050$^{\dagger}$ &SP  \\
 Nov 26, 99 & 51509.430 & 0.652 & 10.867 & 11.530 & 10.388 &  0.289 &SP  \\
 Dec 22, 99 & 51535.364 & 0.686 & 11.078 & 11.620 & 10.532 &  0.437 &SP  \\
 Jan 08, 00 & 51552.321 & 0.709 & 10.984 & 11.684 & 10.610 &  0.487 &SP  \\
 Aug 20, 00 & 51777.449 & 0.005 & 11.590 & 11.933 & 10.777 &  0.752 &SP  \\
 Sep 06, 00 & 51794.368 & 0.028 &  9.645 & 10.485 &  9.783 &    --  & SL \\
 Sep 08, 00 & 51795.531 & 0.029 &  9.691 & 10.438 &  9.748 &  0.106 &SP  \\
 Sep 09, 00 & 51796.555 & 0.031 &  9.689 & 10.383 &  9.691 &
3.512$^{\dagger}$ &SP  \\
 Sep 09, 00 & 51797.363 & 0.032 &  9.711 & 10.396 &  9.695 &    --  & SL \\
 Sep 11, 00 & 51798.612 & 0.033 &  9.635 & 10.202 &  9.532 &
3.406$^{\dagger}$ &SP  \\
 Sep 27, 00 & 51815.331 & 0.055 &  9.658 & 10.054 &  9.436 &
3.318$^{\dagger}$ &SP  \\
 Sep 29, 00 & 51816.510 & 0.057 &  9.555 &  9.952 &  9.350 & -0.186 &SP  \\
 Sep 29, 00 & 51817.263 & 0.058 &  9.562 &  9.970 &  9.367 & -0.176 &SP  \\
 Oct 13, 00 & 51831.276 & 0.076 &  9.094 &  9.524 &  9.013 & --     & SL \\
 Oct 20, 00 & 51838.286 & 0.086 &  9.163 &  9.533 &  9.015 & --     & SL \\
 Oct 21, 00 & 51839.369 & 0.087 &  9.135 &  9.495 &  8.998 & --     & SL \\
 Oct 22, 00 & 51840.475 & 0.088 &  9.131 &  9.455 &  8.991 & -0.387 &SP  \\
 Nov 09, 00 & 51858.360 & 0.112 &  9.106 &  9.533 &  9.064 & -0.347 &SP  \\
 Nov 11, 00 & 51860.249 & 0.114 &  9.029 &  9.610 &  9.122 & --     & SL \\
 Nov 13, 00 & 51862.326 & 0.117 &  9.068 &  9.627 &  9.167 & -0.284 &SP  \\
 Nov 22, 00 & 51871.313 & 0.129 &  8.601 &  9.408 &  8.944 & --     & SL \\
 Nov 28, 00 & 51877.428 & 0.137 &  8.410 &  9.252 &  8.832 & -0.557 &SP  \\
 Dec 01, 00 & 51880.311 & 0.141 &  8.398 &  9.334 &  8.864 & --     & SL \\
 Dec 18, 00 & 51897.240 & 0.163 &  8.364 &  9.278 &  8.850 & --     & SL \\
 Dec 22, 00 & 51901.309 & 0.169 &  8.420 &  9.410 &  8.949 & --     & SL \\
 Dec 24, 00 & 51903.184 & 0.171 &  8.441 &  9.280 &  8.883 & -0.606 &SP  \\
 Dec 24, 00 & 51903.344 & 0.171 &   --   &  9.380 &  8.890 & --     & R  \\
 Dec 27, 00 & 51906.276 & 0.175 &  8.415 &  9.328 &  8.935 & -0.647 &SP  \\
 Jan 12, 01 & 51922.346 & 0.196 &  8.575 &  9.492 &  9.013 & -0.465 &SP  \\
 Jan 15, 01 & 51925.304 & 0.200 &  8.588 &  9.491 &  9.033 & -0.447 &SP  \\
 Jan 19, 01 & 51929.212 & 0.205 &  8.573 &  9.586 &  9.086 & --     & SL \\
 Feb 11, 01 & 51952.236 & 0.236 &  8.749 &  9.761 &  9.234 & --     & SL \\
 Feb 16, 01 & 51957.229 & 0.242 &   --   &   --   &  9.290 & --     & R  \\
 Feb 16, 01 & 51957.276 & 0.242 &  8.797 &  9.725 &  9.207 & -0.331 &SP  \\
 Feb 17, 01 & 51958.228 & 0.244 &   --   &   --   &  9.240 & --     & R  \\
 Mar 06, 01 & 51975.248 & 0.266 &  8.907 &  9.824 &  9.257 & -0.286 &SP  \\
 May 03, 01 & 52032.553 & 0.342 &  8.880 &  9.902 &  9.331 & --     & SL \\
 May 21, 01 & 52050.526 & 0.365 &  9.035 & 10.111 &  9.516 & --     & SL \\
%\hline   
%\end{tabular}
%\end{center}
%\end{table}
%%        
%\addtocounter{table}{-1}
%\begin{table}[!ht]
%\small   
%\begin{center}
%\caption{continued}
%\begin{tabular}{cccccccc}
%\hline   
%Date & JD$_{hel}$ & Phase$^{\star}$ & $U$ & $B$ & $V$ & $\Delta R$ & Obs  \\
%     &-2\,400\,000&                 &     &     &     &            &      \\
%\hline 
 May 25, 01 & 52054.503 & 0.370 &  8.966 & 10.026 &  9.461 & --     & SL \\
 Jul 13, 01 & 52104.462 & 0.436 &  9.166 & 10.254 &  9.601 & --     & SL \\
 Jul 31, 01 & 52122.470 & 0.460 &  9.255 & 10.327 &  9.690 & --     & SL \\
 Aug 16, 01 & 52138.387 & 0.481 &  9.3   &   --   &  9.85 & --      & SL:: \\
 Aug 30, 01 & 52152.472 & 0.500 &  9.257 & 10.254 &  9.610 &
3.366$^{\dagger}$ &SP  \\
 Sep 29, 01 & 52182.479 & 0.539 &  9.302 & 10.188 &  9.591 & -0.066 &SP  \\
 Oct 15, 01 & 52198.281 & 0.560 &  9.370 & 10.467 &  9.794 & --     & SL \\
 Oct 18, 01 & 52201.276 & 0.564 &  9.532 & 10.560 &  9.865 &  0.079 &SP  \\
 Oct 27, 01 & 52210.211 & 0.576 &  9.623 & 10.630 &  9.910 &  0.095 &SP  \\
 Nov 17, 01 & 52231.369 & 0.604 &  9.661 & 10.751 &  9.975 & --     & SL \\
 Nov 18, 01 & 52232.342 & 0.605 &  9.700 &   --   &  9.960 &  0.150 &SP:: \\
 Nov 23, 01 & 52237.255 & 0.611 &  9.74  & 10.83  & 10.03  & --     &SL:: \\
\hline    
\end{tabular}
\end{center}
$^{\star}$ $JD_{\rm sp. conj.} = 2\,445\,703.0 + 758.8\times E$ 
           (Mikolajewska \& Kenyon 1996) \\
$^{\dagger}$ $\Delta R$ = Z\,And - SAO\,35642
\vspace*{-5mm}
\normalsize
\end{table}

\subsection{BF\,Cyg}

The photometric measurements of BF\,Cyg (MWC\,315, Hen\,1747) are 
given in Table 3. Stars 
HD\,183650 (SAO\,68384, $V$=6.96, $B-V$=0.71, $U-B$=0.34)
and BD+30\,3594 (LF2+3011, UBV\,16553, $V$=9.54, $B-V$=1.20, $U-B$=1.70)
were used as the comparison and check, respectively. 
In addition, for our CCD photometry we used the following sequence 
of standard stars:
SAO\,87137, BD+29\,\-35\-81 and GSC\,02137-02086. 
Figure 3 shows the $U,B,V$ LCs from 1995.8. Periodic wave-like 
variation in the optical continuum reflects a quiescent 
phase of this star. The profile of LCs is not a simple 
sinusoid through the orbital period, but reflects rather 
complex shape and variation of the nebula in the binary. 
For example, the star's brightness at the light minima 
gradually decreases. The JD~2\,450\,684.5 minimum occurred at 
$U\sim 12.5$, $B\sim 12.9$, $V\sim 12.4$, while the recent minimum 
observed around JD~2\,452\,173 showed magnitudes of 
$U\sim 12.8$, $B\sim 13.3$, $V\sim 12.7$.
In addition, short-term unpredictable fading in the star's brightness, 
in all bands was observed at $\sim$JD~2\,451\,089 ($\varphi \sim 0.55$) 
and at $\sim$JD~2\,451\,700 ($\varphi \sim 0.38$). This event could be 
ascribed to an occultation of the hot component by an extra blob of 
the neutral material from the giant component. 
Finally, a complex secondary minima can be recognized in $U,B,V$ LCs 
around 1998.6 and 2000.7 (see Fig. 3), when the hot star is in front 
($\varphi \sim 0.5$). Such evolution reflects a shrinking of the nebula 
that is probably due to a decrease of the ionizing photons producing 
by the hot star. 
%
%==============================|
%        Fig. 3: BF Cyg        |
%==============================|
%%\vspace*{0.25cm}
\begin{figure}[!ht]
  \centering
  \centerline{\hbox{
  \psfig{figure=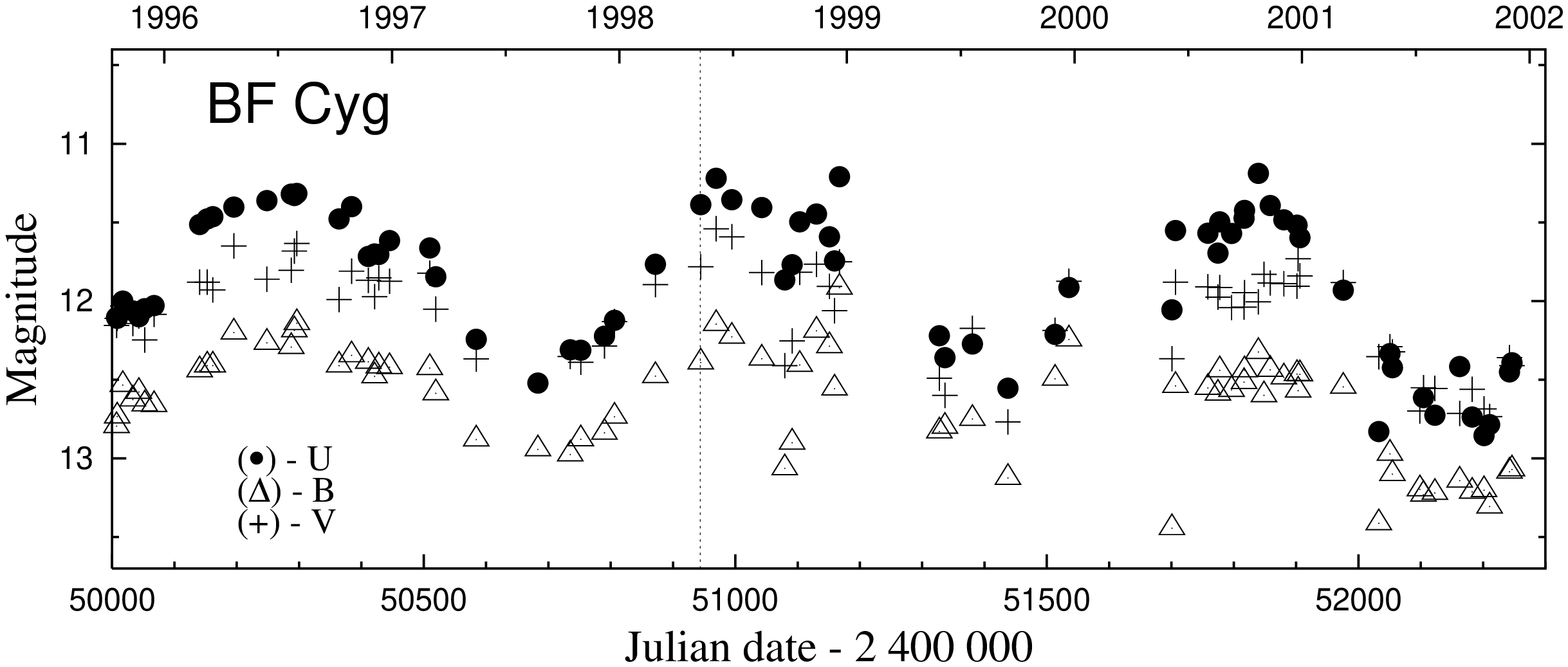,width=12cm,angle=-90}}}
%%\vspace{0.25cm}
\caption[ ]{The $UBV$ LCs of BF\,Cyg. New data presented in this paper 
are plotted from the vertical dotted line. 
}
%\end{figure}
%
%==============================|
%        Fig. 4: CH Cyg        |
%==============================|
%\vspace*{0.25cm}
%\begin{figure}[!ht]
  \centering
  \centerline{\hbox{
  \psfig{figure=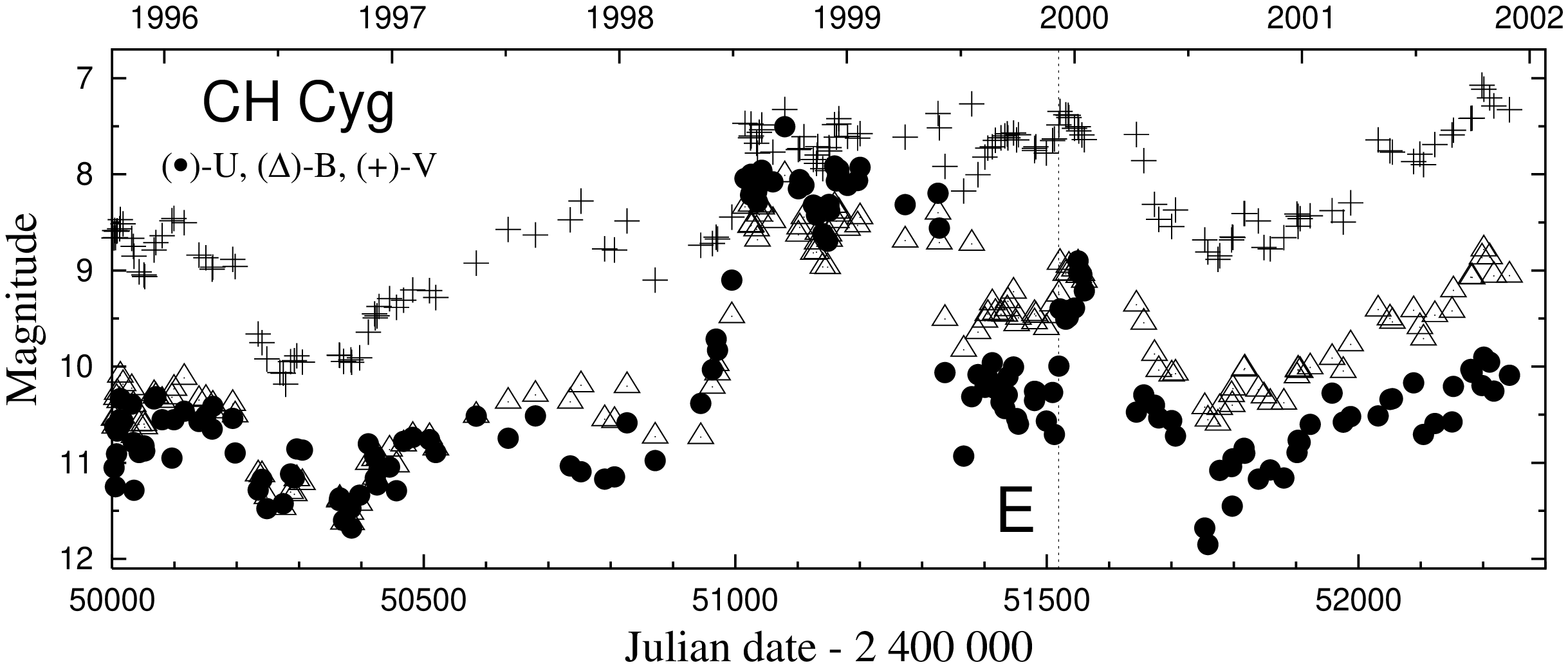,width=12cm,angle=-90}}}
%\vspace{0.25cm}
\caption[ ]{As in Fig. 3, but for CH\,Cyg. {\bf E} denotes the eclipse 
in the outer binary. 
}
\end{figure}
%
%
%==============================|
%       Table 3: BF Cyg        |
%==============================|
%
\begin{table}[!ht]
%\small
\scriptsize
\begin{center}
\caption{$U,B,V,R$ observations of BF\,Cyg}
\begin{tabular}{lccccccc}
\hline
Date & JD~24... & Phase$^{\star}$ & $U$ & $B$ & $V$ & $\Delta R$  & Obs  \\
\hline
 May 11, 98 & 50944.508 & 0.378 & 11.387 & 12.391 & 11.781 &  3.960 &SP  \\
 Jun 04, 98 & 50969.458 & 0.411 & 11.219 & 12.147 & 11.540 &  3.830 &SP  \\
 Jun 29, 98 & 50994.428 & 0.444 & 11.356 & 12.228 & 11.592 &  3.847 &SP  \\
 Aug 16, 98 & 51042.484 & 0.508 & 11.406 & 12.366 & 11.819 &  4.054 &SP  \\
 Sep 22, 98 & 51079.341 & 0.556 & 11.866 & 13.064 & 12.410 &  4.286 &SP  \\
 Oct 04, 98 & 51091.252 & 0.572 & 11.769 & 12.900 & 12.253 &  4.190 &SP  \\
 Oct 16, 98 & 51103.281 & 0.588 & 11.496 & 12.406 & 11.817 &  4.124 &SP  \\
 Nov 12, 98 & 51130.278 & 0.624 & 11.447 & 12.189 & 11.766 &  4.096 &SP  \\
 Dec 03, 98 & 51151.201 & 0.651 & 11.592 & 12.287 & 11.906 &  4.254 &SP  \\
 Dec 11, 98 & 51159.228 & 0.662 & 11.744 & 12.557 & 12.061 &  4.316 &SP  \\
 Dec 19, 98 & 51167.322 & 0.672 & 11.210 & 11.915 & 11.750 &  4.300 &SP  \\
 May 29, 99 & 51327.533 & 0.884 & 12.220 & 12.830 & 12.490 &  4.770 &SP  \\
 Jun 07, 99 & 51336.500 & 0.896 & 12.36 & 12.8 & 12.6 & 3.38$^\dagger$&SP::\\
 Jul 21, 99 & 51380.507 & 0.954 & 12.273 & 12.750 & 12.174 &  4.605 &SP  \\
 Sep 15, 99 & 51437.353 & 0.029 & 12.554 & 13.125 & 12.768 &  4.994 &SP  \\
 Nov 30, 99 & 51513.273 & 0.129 & 12.212 & 12.494 & 12.187 &2.451$^\dagger$ &SP\\
 Dec 22, 99 & 51535.205 & 0.158 & 11.913 & 12.244 & 11.874 &  4.478 &SP  \\
 Jun 04, 00 & 51700.487 & 0.376 & 12.055 & 13.445 & 12.367 &  4.131 &SP  \\
 Jun 10, 00 & 51706.488 & 0.384 & 11.551 & 12.540 & 11.880 & --     & SL \\
 Aug 01, 00 & 51758.388 & 0.453 & 11.569 & 12.552 & 11.909 & --     & SL \\
 Aug 17, 00 & 51774.372 & 0.474 & 11.695 & 12.589 & 11.975 & --     & SL \\
 Aug 20, 00 & 51777.369 & 0.478 & 11.496 & 12.441 & 11.914 &  4.094 &SP  \\
 Sep 08, 00 & 51796.461 & 0.503 & 11.570 & 12.566 & 12.041 &  4.172 &SP  \\
 Sep 28, 00 & 51816.455 & 0.530 & 11.474 & 12.516 & 12.038 &  4.175 &SP  \\
 Sep 29, 00 & 51817.299 & 0.531 & 11.424 & 12.440 & 11.946 &  4.133 &SP  \\
 Oct 21, 00 & 51839.293 & 0.560 & 11.188 & 12.324 & 12.004 & --     & SL \\
 Oct 30, 00 & 51848.233 & 0.571 &  --    & 12.60  & 11.83  & --     & R  \\
 Nov 09, 00 & 51858.230 & 0.585 & 11.392 & 12.437 & 11.885 &  4.076 &SP  \\
 Dec 01, 00 & 51880.236 & 0.614 & 11.484 & 12.486 & 11.890 & --     & SL \\
 Dec 22, 00 & 51901.193 & 0.641 & 11.517 & 12.464 & 11.905 & --     & SL \\
 Dec 24, 00 & 51903.188 & 0.644 &  --    & 12.57  & 11.73  & --     & R  \\ 
 Dec 27, 00 & 51906.191 & 0.648 & 11.597 & 12.466 & 11.840 &  4.020 &SP  \\
 Mar 07, 01 & 51975.614 & 0.740 & 11.931 & 12.545 & 11.882 &  4.111 &SP  \\
 May 02, 01 & 52032.491 & 0.815 & 12.830 & 13.413 & 12.354 & --     & SL \\
 May 20, 01 & 52050.431 & 0.839 & 12.334 & 12.972 & 12.290 & --     & SL \\
 May 24, 01 & 52054.425 & 0.844 & 12.423 & 13.101 & 12.322 & --     & SL \\
 Jul 07, 01 & 52098.436 & 0.902 &  --    & 13.198 & 12.699 &  4.384 &SP::  \\
 Jul 13, 01 & 52104.431 & 0.910 & 12.614 & 13.231 & 12.551 & --     & SL \\
 Jul 31, 01 & 52122.411 & 0.934 & 12.726 & 13.220 & 12.556 & --     & SL \\
 Sep 09, 01 & 52162.288 & 0.986 & 12.417 & 13.142 & 12.715 & --     & SL \\
 Sep 29, 01 & 52182.419 & 0.013 & 12.737 & 13.213 & 12.561 &  4.985 &SP  \\
 Oct 18, 01 & 52201.335 & 0.038 & 12.855 & 13.203 & 12.686 &  4.994 &SP  \\
 Oct 27, 01 & 52210.332 & 0.050 & 12.786 & 13.306 & 12.735 &  5.007 &SP  \\
 Nov 28, 01 & 52242.257 & 0.092 & 12.450 & 13.085 & 12.360 &  4.751 &SP  \\
\hline    
\end{tabular}
\end{center}
$^{\star}$ $Min = JD\,2\,415\,065 + 757.3\times E$ 
           (Pucinskas 1970) \\
$^{\dagger}$ $\Delta R$ = BF\,Cyg - BD+30\,3594
\normalsize
\end{table}

\subsection{CH\,Cyg}

Our new photometry of CH\,Cyg (HD\,182917, BD+49\,2999) is listed in 
Table 4. Measurements were done with respect to the same standard 
stars as in the Paper~IX (Skopal et al. 2000a). Additional comparison 
stars, SAO\,48428, SAO\,31628 and BD+49\,3005, were used for our 
CCD photometry. 
Figure 4 displays the $U,B,V$ LCs covering the period from 1995.8. 
It shows that the recent episode of activity ended in Spring of 2000. 
A more detailed analysis of this active phase is given by 
Eyres et al. (2002). Our observations allowed us 
to determine position of a broad minimum at JD~2\,451\,426$\pm 3$, 
caused by the eclipse of the active star in the inner binary 
(the symbiotic pair) by the cool giant on the outer orbit. Profile 
of the $U,B,V$ LCs observed after the recent activity, around 2000.7, 
is similar to that occurred after the previous active phase, 
around 1996.5. Recent observations indicate a slow increase in 
the star's brightness. 
%
%==============================|
%       Table 4: CH Cyg        |
%==============================|
%
\begin{table}[!ht]
%\vspace*{8mm}
%\small
\scriptsize
\begin{center}
\caption{$U,~B,~V,~R$ observations of CH\,Cyg}
\begin{tabular}{lccccccc}
\hline
Date & JD~24... & Phase$^{\star}$ & $U$ & $B$ & $V$ & $\Delta R$  & Obs  \\
\hline
 Dec 08, 99 & 51521.168 & 0.451 &  9.400 &  8.921 &  7.346 & --     & SL \\
 Dec 17, 99 & 51530.181 & 0.463 &  9.507 &  9.036 &  7.380 & --     & SL \\
 Dec 21, 99 & 51534.175 & 0.468 &  9.486 &  9.034 &  7.385 & --     & SL \\
 Dec 22, 99 & 51535.261 & 0.470 &  9.414 &  8.952 &  7.408 & -0.951 &SP  \\
 Dec 31, 99 & 51544.222 & 0.482 &  9.391 &  8.992 &  7.520 & --     & SP \\
 Jan 06, 00 & 51550.190 & 0.490 &  8.900 &  8.929 &  7.503 & --     & SL \\
 Jan 07, 00 & 51551.186 & 0.491 &  9.023 &  9.058 &  7.547 & --     & SL \\
 Jan 12, 00 & 51556.184 & 0.498 &  9.045 &  9.037 &  7.590 & --     & SL \\
 Jan 16, 00 & 51560.202 & 0.503 &  9.212 &  9.108 &  7.639 & --     & SL \\
 Apr 09, 00 & 51643.566 & 0.613 & 10.473 &  9.357 &  7.586 & --     & SL \\
 Apr 20, 00 & 51655.386 & 0.629 & 10.294 &  9.545 &  7.858 & --     & SL \\
 May 08, 00 & 51672.509 & 0.651 & 10.402 &  9.869 &  8.313 & -0.257 &SP  \\
 May 15, 00 & 51679.534 & 0.661 & 10.534 & 10.036 &  8.469 & -0.114 &SP  \\
 Jun 04, 00 & 51700.415 & 0.688 & 10.562 & 10.084 &  8.543 & -0.057 &SP  \\
 Jun 10, 00 & 51706.404 & 0.696 & 10.721 & 10.073 &  8.370 & --     & SL \\
 Jul 27, 00 & 51753.381 & 0.758 & 11.680 & 10.422 &  8.683 & --     & SL \\
 Aug 01, 00 & 51758.335 & 0.765 & 11.852 & 10.552 &  8.808 & --     & SL \\
 Aug 17, 00 & 51774.322 & 0.786 & 12.420 & 10.595 &  8.885 & --     & SL \\
 Aug 20, 00 & 51777.411 & 0.790 & 11.079 & 10.431 &  8.847 &  0.219 &SP  \\
 Sep 08, 00 & 51796.408 & 0.815 & 11.040 & 10.297 &  8.683 &  0.086 &SP  \\
 Sep 09, 00 & 51797.274 & 0.816 & 11.452 & 10.401 &  8.651 & --     & SL \\
 Sep 11, 00 & 51798.576 & 0.818 & 10.960 & 10.230 &  8.656 &  0.034 &SP  \\
 Sep 28, 00 & 51816.407 & 0.842 & 10.850 & 10.031 &  8.408 & -0.159 &SP  \\
 Sep 29, 00 & 51817.344 & 0.843 & 10.900 & 10.020 &  8.407 & -0.160 &SP  \\
 Oct 21, 00 & 51839.253 & 0.872 & 11.170 & 10.240 &  8.485 & --     & SL \\
 Oct 30, 00 & 51848.257 & 0.884 &  --    & 10.31  &  8.76  & --     & R  \\
 Nov 09, 00 & 51858.288 & 0.897 & 11.076 & 10.379 &  8.774 &  0.089 &SP  \\
 Dec 01, 00 & 51880.195 & 0.926 & 11.159 & 10.375 &  8.658 & --     & SL \\
 Dec 22, 00 & 51901.230 & 0.954 & 10.894 & 10.101 &  8.413 & --     & SL \\
 Dec 24, 00 & 51903.205 & 0.957 &  --    & 10.04  &  8.54  & --     & R  \\
 Dec 24, 00 & 51903.256 & 0.957 & 10.767 & 10.032 &  8.436 & -0.132 &SP  \\
 Dec 27, 00 & 51906.239 & 0.961 & 10.787 & 10.046 &  8.462 & -0.113 &SP  \\
 Jan 13, 01 & 51922.682 & 0.982 & 10.602 & 10.009 &  8.431 & -0.196 &SP  \\
 Feb 17, 01 & 51957.571 & 0.029 & 10.275 &  9.912 &  8.378 & -0.225 &SP  \\
 Mar 07, 01 & 51975.554 & 0.052 & 10.579 & 10.047 &  8.497 & -0.143 &SP  \\
 Mar 19, 01 & 51987.502 & 0.068 & 10.519 &  9.769 &  8.297 & -0.249 &SP  \\
 May 01, 01 & 52031.431 & 0.126 & 10.513 &  9.406 &  7.641 & --     & SL \\
 May 20, 01 & 52050.387 & 0.151 & 10.342 &  9.501 &  7.756 & --     & SL \\
 May 24, 01 & 52054.384 & 0.157 & 10.334 &  9.534 &  7.770 & --     & SL \\
 Jun 27, 01 & 52088.417 & 0.202 & 10.17  &  9.42  &  7.87  & -0.72  &SP: \\
 Jul 07, 01 & 52098.375 & 0.215 &  --    &  9.591 &  7.789 & -0.661 &SP  \\
 Jul 13, 01 & 52104.350 & 0.223 & 10.706 &  9.708 &  7.899 & --     & SL \\
 Jul 31, 01 & 52122.377 & 0.247 & 10.591 &  9.465 &  7.690 & --     & SL \\
 Aug 28, 01 & 52150.314 & 0.283 & 10.575 &  9.420 &  7.590 & --     & SL \\
 Aug 30, 01 & 52152.372 & 0.286 & 10.207 &  9.207 &  7.540 & -0.886 &SP  \\
 Sep 27, 01 & 52180.373 & 0.323 & 10.029 &  9.070 &  7.417 & -1.005 &SP  \\
 Sep 29, 01 & 52182.371 & 0.326 & 10.060 &  9.073 &  7.416 & -0.998 &SP  \\
 Oct 15, 01 & 52198.234 & 0.347 & 10.197 &  8.866 &  7.071 & --     & SL \\
 Oct 18, 01 & 52201.231 & 0.351 &  9.902 &  8.783 &  7.113 & -1.223 &SP  \\
 Oct 27, 01 & 52210.332 & 0.363 &  9.953 &  8.864 &  7.205 & -1.169 &SP  \\
 Nov 03, 01 & 52217.275 & 0.372 & 10.256 &  9.055 &  7.287 & --     & SL \\
 Nov 28, 01 & 52242.308 & 0.405 & 10.091 &  9.053 &  7.328 & -1.006 &SP  \\
\hline
\end{tabular}
\end{center}
$^{\star}$ $Min = JD\,2\,445\,888 + 756\times E$
           (Skopal 1995)
%
%\vspace*{15mm}
%
\normalsize
\end{table}

\subsection{V1329\,Cyg}

Our observations of the symbiotic nova V1329\,Cyg (HBV\,475) are
given in Table 5. The stars 
BD+35\,4290 ($V$=10.34, $B-V$=1.07, $U-B$=0.88) and 
BD+35\,4294 ($V$=10.16, $B-V$=1.07) 
were used as the comparison and check, respectively.
Figure 5 shows the $U,B,V$ LCs from 1987.5. 
A wave-like variation, which mimics a reflection effect, represents 
a dominant feature of these LCs. At light maxima the star's brightness 
is highest in the $U$ band, while at the minima, $U\,\sim\,V$. 
This indicates a strong nebular component of radiation in the system 
resulting from both a high luminosity of the hot ionizing object and 
a high mass loss rate from the cool component (M6\,III giant). 
%
%==============================|
%      Fig. 5: V1329 Cyg       |
%==============================|
%\vspace*{0.25cm}
\begin{figure}[!ht]
  \centering
  \centerline{\hbox{
  \psfig{figure=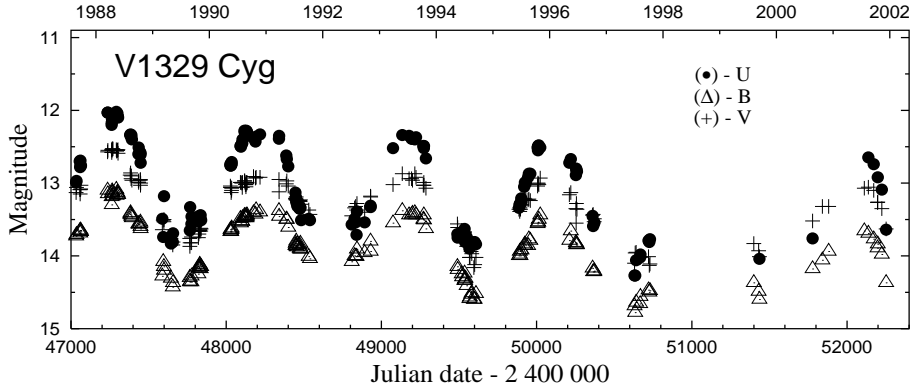,width=12cm,angle=-90}}}
%\vspace{0.25cm}
\caption[ ]{As in Fig. 3, but for V1329\,Cyg. Our data (from 1999 August 9) 
are compiled with those of Chochol et al. (1999).
}
\end{figure}
%
%
%==============================|
%      Table 5: V1329 Cyg      |
%==============================|
%
\begin{table}[!ht]
%\small
\scriptsize
\begin{center}
\caption{$U,B,V$ observations of V1329\,Cyg.} 
\begin{tabular}{lcccccc}
\hline
    Date  & JD~24... & Phase$^{\star}$ & $U$ & $B$ & $V$      & Obs\\
\hline
 Aug 09, 99 & 51399.566 & 0.752 &  --    & 14.370 & 13.830    & SL \\
 Sep 08, 99 & 51430.468 & 0.784 &  --    & 14.490 & 13.930    & SL \\
 Sep 13, 99 & 51435.410 & 0.790 & 14.040 & 14.600 & 14.010    & SL \\
 Aug 21, 00 & 51778.417 & 0.148 & 13.760 & 14.180 & 13.520    & SL \\
 Oct 23, 00 & 51841.249 & 0.213 &  --    & 14.060 & 13.320    & SL \\
 Oct 30, 00 & 51848.278 & 0.221 &  --    & 13.96  & 13.13     & R  \\
 Dec 02, 00 & 51881.193 & 0.255 &  --    & 13.940 & 13.320    & SL \\
 Jul 21, 01 & 52111.518 & 0.495 &  --    & 13.660 & 13.070    & SL \\
 Aug 15, 01 & 52137.392 & 0.522 & 12.647 & 13.692 & 13.055    & SL \\
 Sep 19, 01 & 52172.313 & 0.559 & 12.740 & 13.770 & 13.100    & SL \\
 Oct 13, 01 & 52196.279 & 0.584 &  --    & 13.894 & 13.266    & SL \\
 Oct 15, 01 & 52198.326 & 0.586 & 12.920 & 13.840 & 13.260    & SL \\
 Nov 10, 01 & 52224.322 & 0.613 & 13.090 & 13.980 & 13.350    & SL \\
 Dec 08, 01 & 52252.191 & 0.642 & 13.64  & 14.37  & 13.63     & SL::\\
\hline
\end{tabular}
\end{center}
$^{\star}$ $JD_{\rm eclipse} = 2\,427\,687 + 958.0 \times E$
           (Schild \& Schmid 1997)
\normalsize 
\end{table}
%
%
%
%==============================|
%       Table 6: AG Dra        |
%==============================|
%
\begin{table}[!ht]
%\small
\scriptsize
\begin{center}
\caption{$U,B,V,R$ observations of AG\,Dra}
\begin{tabular}{lccccccc}
\hline
Date & JD~24... & Phase$^{\star}$ & $U$ & $B$ & $V$ & $\Delta R$  & Obs  \\
\hline
 May 29, 98 & 50963.380 & 0.579 & 11.102 & 11.113 &  9.767 & --     & SL \\
 Jun 06, 98 & 50971.366 & 0.593 & 11.062 & 11.150 &  9.795 & --     & SL \\
 Jun 29, 98 & 50994.468 & 0.635 & 10.800 & 11.052 &  9.755 & -0.651 &SP  \\
 Jul 20, 98 & 51015.372 & 0.673 & 10.086 & 10.721 &  9.631 & --     & SL \\
 Jul 21, 98 & 51016.361 & 0.675 &  9.531 & 10.427 &  9.461 & -0.868 &SP  \\
 Jul 29, 98 & 51024.362 & 0.689 &  9.107 & 10.169 &  9.291 & --     & SL \\
 Jul 30, 98 & 51025.355 & 0.691 &  9.111 & 10.231 &  9.316 & --     & SL \\
 Aug 05, 98 & 51031.331 & 0.702 &  9.3   &  --    &   --   & --    & SP::\\
 Aug 09, 98 & 51034.524 & 0.708 &  9.264 & 10.244 &  9.338 & -1.021 &SP  \\
 Aug 15, 98 & 51041.328 & 0.720 &  9.086 & 10.296 &  9.394 & --     & SL \\
 Aug 17, 98 & 51042.563 & 0.723 &  9.097 & 10.227 &  9.346 & -1.055 &SP  \\
 Sep 04, 98 & 51060.553 & 0.755 &  9.473 & 10.514 &  9.531 & -0.926 &SP  \\
 Sep 23, 98 & 51079.593 & 0.790 &  9.834 & 10.617 &  9.546 & -0.906 &SP  \\
 Oct 14, 98 & 51101.290 & 0.829 & 10.350 & 10.881 &  9.677 & --     & SL \\
 Oct 23, 98 & 51110.245 & 0.846 & 10.444 & 10.878 &  9.683 & --     & SL \\
 Nov 07, 98 & 51125.255 & 0.873 & 10.727 & 10.914 &  9.712 & --     & SL \\
 Nov 12, 98 & 51130.478 & 0.882 & 10.815 & 10.925 &  9.684 & -0.729 &SP  \\
 Dec 12, 98 & 51159.657 & 0.936 & 11.097 & 10.989 &  9.699 & -0.678 &SP  \\
 Jan 06, 99 & 51185.389 & 0.982 & 11.230 & 11.060 &  9.774 & -0.644 &SP  \\
 Jan 21, 99 & 51200.278 & 0.009 & 11.277 & 11.036 &  9.740 & --     & SL \\
 Jan 24, 99 & 51202.566 & 0.014 & 11.294 & 11.000 &  9.749 & -0.643 &SP  \\
 Feb 27, 99 & 51237.330 & 0.077 & 11.282 & 11.081 &  9.747 & --     & SL \\
 Apr 03, 99 & 51272.473 & 0.141 & 11.347 & 11.161 &  9.815 & --     & SL \\
 May 28, 99 & 51327.463 & 0.241 & 11.406 & 11.144 &  9.746 & -0.663 &SP  \\
 Sep 14, 99 & 51436.311 & 0.439 & 11.316 & 11.108 &  9.773 & --     & SL \\
 Sep 15, 99 & 51437.290 & 0.441 & 11.332 & 11.116 &  9.767 & -0.635 &SP  \\
 Sep 25, 99 & 51446.609 & 0.458 & 11.462 & 11.154 &  9.822 & -0.592 &SP  \\
 Nov 27, 99 & 51509.658 & 0.572 & 11.190 & 11.048 &  9.707 & -0.682 &SP  \\
 Nov 28, 99 & 51510.606 & 0.574 & 11.163 & 11.085 &  9.727 & -0.654 &SP  \\
 Dec 23, 99 & 51535.625 & 0.619 & 11.174 & 11.029 &  9.645 & -0.730 &SP  \\
 Jan 09, 00 & 51552.515 & 0.650 & 11.096 & 11.072 &  9.743 & -0.644 &SP  \\
 Feb 22, 00 & 51597.472 & 0.732 & 11.027 & 11.038 &  9.721 & -0.660 &SP  \\
 Apr 20, 00 & 51655.344 & 0.837 &  --     & 11.028 &  9.764 & --     & SL \\
 Jun 04, 00 & 51700.372 & 0.919 & 11.062 & 11.146 &  9.790 & -0.606 &SP  \\
 Aug 17, 00 & 51774.415 & 0.054 & 11.161 & 11.109 &  9.779 & --     & SL \\
 Aug 21, 00 & 51777.566 & 0.060 &  --    & 11.22  &   9.92 &  -0.47 &SP: \\
 Sep 09, 00 & 51797.298 & 0.095 & 11.174 & 11.069 &  9.722 & --     & SL \\
 Sep 28, 00 & 51816.302 & 0.130 & 11.292 & 11.136 &  9.758 & -0.606 &SP  \\
 Oct 21, 00 & 51839.226 & 0.172 & 11.528 & 11.109 &  9.732 & --     & SL \\
 Oct 30, 00 & 51848.200 & 0.188 &  --    & 11.03  &  9.80  & --     & R  \\
 Dec 22, 00 & 51900.694 & 0.284 & 11.657 & 11.162 &  9.772 & --     & SL \\
 Dec 24, 00 & 51903.354 & 0.288 &  --    &  --    &  9.703 & -0.846 &SP  \\
 Dec 25, 00 & 51904.638 & 0.291 &  --    & 11.07  &  9.86  & --     & R  \\ 
 Jan 10, 01 & 51919.652 & 0.318 & 11.703 & 11.161 &  9.776 & --     & SL \\
 Jan 13, 01 & 51922.616 & 0.323 & 11.684 & 11.186 &  9.775 & -0.600 &SP  \\
 Feb 16, 01 & 51957.401 & 0.387 & 11.560 & 11.147 &  9.780 & -0.560 &SP  \\
 Feb 17, 01 & 51958.540 & 0.389 &  --    & 11.10  &  9.91  & --     &R   \\
 Mar 06, 01 & 51975.495 & 0.420 & 11.501 & 11.171 &  9.794 & -0.607 &SP  \\
 Mar 18, 01 & 51987.446 & 0.441 & 11.442 & 11.182 &  9.799 & -0.574 &SP  \\
 Mar 18, 01 & 51987.411 & 0.441 & 11.360 & 11.129 &  9.778 & --     & SL \\
 May 01, 01 & 52031.400 & 0.521 & 11.133 & 11.036 &  9.704 & --     & SL \\
 May 20, 01 & 52050.352 & 0.556 & 11.101 & 11.040 &  9.699 & --     & SL \\
 Jul 13, 01 & 52104.386 & 0.654 & 10.804 & 11.012 &  9.693 & --     & SL \\
 Aug 01, 01 & 52122.535 & 0.687 & 10.548 & 10.964 &  9.704 & --     & SL \\
 Aug 30, 01 & 52152.422 & 0.741 & 10.560 & 11.071 &  9.823 & -0.606 &SP  \\
 Sep 20, 01 & 52173.268 & 0.779 & 10.571 & 10.988 &  9.735 & --     & SL \\
 Sep 27, 01 & 52180.314 & 0.792 & 10.456 & 10.863 &  9.674 & -0.691 &SP  \\
 Sep 29, 01 & 52182.313 & 0.796 & 10.429 & 10.916 &  9.695 & -0.691 &SP  \\
 Oct 16, 01 & 52199.234 & 0.827 & 10.147 & 10.745 &  9.609 & --     & SL \\
\hline   
\end{tabular}
\end{center}
\end{table}
\addtocounter{table}{-1}
\begin{table}[!ht]
\small   
\begin{center}
\caption{continued}
\begin{tabular}{cccccccc}
\hline   
Date & JD$_{hel}$ & Phase$^{\star}$ & $U$ & $B$ & $V$ & $\Delta R$ & Obs  \\
     &-2\,400\,000&                 &     &     &     &            &      \\
\hline 
 Oct 18, 01 & 52201.441 & 0.831 & 10.309 & 10.779 &  9.623 & -0.749 &SP  \\
 Oct 27, 01 & 52210.455 & 0.847 &  9.93  &  --    &   --   & --     & SP:: \\
 Oct 31, 01 & 52214.210 & 0.854 &  9.877 & 10.632 &  9.540 & --     & SL \\
 Nov 03, 01 & 52217.244 & 0.859 &  9.830 & 10.619 &  9.533 & --     & SL \\
 Nov 23, 01 & 52237.197 & 0.896 & 10.117 & 10.744 &  9.580 & --     & SL \\
 Dec 02, 01 & 52246.277 & 0.912 & 10.291 & 10.753 &  9.625 & --     & SL \\
\hline
\end{tabular}
\end{center}
$^{\star}$ $Min = JD\,2\,447\,896.71 + 549.73\times E$
           (G\'alis et al. 1999)
\normalsize     
\end{table}

\subsection{AG\,Dra}

Observations of AG\,Dra (SAO\,16931, BD+67\,922) is compiled in Table 6. 
We used the same standard stars as in the Paper~VIII (Skopal 1998a). 
For the CCD measurements we used an additional standard star, 
GSC\,04195-00369. 
Our measurements cover the recent two eruptions with maxima at 1998.6 
and 2001.8. 
Figure 6 shows the $U,B,V$ LCs covering the recent active phase 
of AG\,Dra, which began in 1994.5 displaying a massive eruption. 
Our data in the figure are compiled with those published by 
Montagni et al. (1996), Greiner et al. (1997), Mikolajewski (1997), 
Tomova \& Tomov (1998), Petr\'{\i}k et al. (1998) and 
Tomov \& Tomova, (2000). Agreement between all data sets is very good. 
All eruptions are most pronounced in the $U$ band. For example, 
for the recent 2000.8 small eruption $\Delta U \sim 1.2$, 
$\Delta B \sim 0.5$ and $\Delta V \sim 0.2$\,mag.
Such behaviour suggests nebular origin of the light during the 
activity. This is also supported by the SED presented by 
Greiner et al. (1997), and the fact that the yellow cool giant 
(spectrum K2) in the system dominates the $V$ band. 
%
%==============================|
%        Fig. 6: AG Dra        |
%==============================|
%\vspace*{0.25cm}
\begin{figure}[!ht]
  \centering
  \centerline{\hbox{
  \psfig{figure=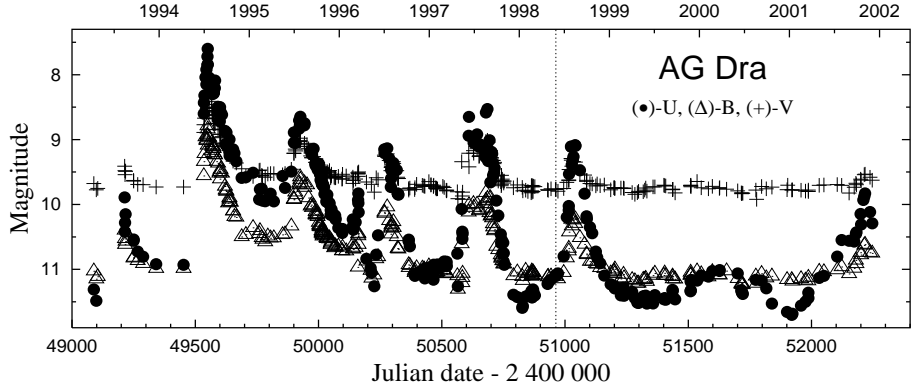,width=12cm,angle=-90}}}
%\vspace{0.25cm}
\caption[ ]{As in Fig. 3, but for AG\,Dra.
}
\end{figure}

\subsection{RW\,Hya}

$U,B,V$ measurements of RW\,Hya (HD 117970) are listed in Table 7. 
Observation was carried out at the San Pedro Observatory during June 
2001 (Sect. 2). 
Stars HD\,118102 (CD-24 10984;  $V$ = 8.94, $B-V$ = 0.53, $U-B$ = 0.11) 
and HD\,117971 (CD-25\,9879; $V$ = 9.69, $B-V$ = 0.44, $U-B$ = -0.03), 
were used as a comparison and a check star, respectively. 
Our photoelectric observations are in very good agreement with those 
published by Munari et al. (1992). This star was also observed visualy 
during a period of 1989.0 to 2001.7. 

Figure 7 shows a phase diagram of our visual observations. It displays 
a complex wave-like variation throughout the orbital cycle. 
The data were folded according to ephemeris of the inferior spectroscopic 
conjunction of the giant star in RW\,Hya (Schild et al. 1996). 
We can see that the light minimum precedes 
the time of the spectroscopic conjunction of the giant (given in 
the figure by the orbital phase 0 or 1). This is consistent with 
revealing of Skopal (1998b) that during quiescent phases the light 
minima occur prior to the time of the inferior conjunction of the cool 
component. The effect is best seen in eclipsing systems. 
%
%
%==============================|
%        Fig. 7: RW Hya        |
%==============================|
%\vspace*{0.25cm}
\begin{figure}[!ht]
  \centering
  \centerline{\hbox{
  \psfig{figure=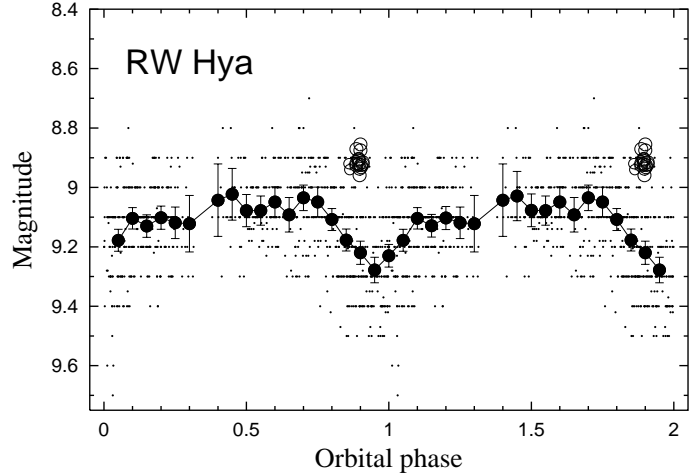,width=9cm,angle=-90}}}
%\vspace{0.25cm}
\caption[ ]{Phase diagram for our visual estimates of RW\,Hya: 
(dots) - all individual observations, ($\bullet$) - maen values made 
in bins of $\Delta \varphi$ = 0.05 with corresponding uncertainty (rms). 
An average uncertainty of one estimate was adopted to 0.3\,mag. 
Compared are our photoelectric $V$ measurements ($\bigcirc$) and those 
of Munari et al. (1992) ($\circ$). Difference between visual and 
photoelectric data sets is probably caused by different comparison 
stars used. 
}
\end{figure}
%
%
%==============================|
%       Table 7: RW Hya        |
%==============================|
%
\begin{table}[!ht]
%\small
\scriptsize
\begin{center}
\caption{$U,B,V$ observations of RW\,Hya}
\begin{tabular}{lcccccc}
\hline
Date & JD~24... & Phase$^{\star}$ & $U$ &   $B$   &   $V$  & Obs \\
\hline
 Jun 02, 01 & 52062.684 & 0.886 & 10.907 & 10.362 &  8.926 &  M  \\
   ------   & 52062.685 & 0.886 & 10.934 & 10.364 &  8.919 &  M  \\
   ------   & 52062.695 & 0.886 & 10.684 & 10.267 &  8.871 &  M  \\
 Jun 05, 01 & 52065.664 & 0.894 & 10.995 & 10.413 &  8.922 &  M  \\
   ------   & 52065.669 & 0.894 & 10.926 & 10.389 &  8.910 &  M  \\
   ------   & 52065.674 & 0.894 & 10.930 & 10.369 &  8.904 &  M  \\
   ------   & 52065.674 & 0.894 & 10.982 & 10.366 &  8.906 &  M  \\
 Jun 06, 01 & 52066.671 & 0.897 & 10.975 & 10.420 &  8.959 &  M  \\
   ------   & 52066.677 & 0.897 & 10.995 & 10.421 &  8.937 &  M  \\
 Jun 07, 01 & 52067.687 & 0.900 & 10.856 & 10.309 &  8.875 &  M  \\
   ------   & 52067.690 & 0.900 & 10.771 & 10.270 &  8.855 &  M  \\
 Jun 08, 01 & 52068.675 & 0.902 & 11.009 & 10.420 &  8.932 &  M  \\
   ------   & 52068.680 & 0.902 & 11.012 & 10.428 &  8.925 &  M  \\
 Jun 10, 01 & 52070.700 & 0.908 & 11.060 & 10.412 &  8.922 &  M  \\
   ------   & 52070.705 & 0.908 & 11.032 & 10.404 &  8.915 &  M  \\
\hline
\end{tabular}
\end{center}
$^{\star}$ $JD_{\rm sp. conj.} = 2\,449\,512 + 370.4\times E$
           (Schild, M\"urset \& Schmutz 1996)
\normalsize     
\end{table}

\subsection{AX\,Per}

The recent measurements of AX\,Per (MWC\,411, GCRV\,896) in 
the $U,B,V,R$ bands are given in Table 8 and showed in Fig. 8. 
Also in this case we used the same standard stars as in Paper~IX. 
For our CCD photometry the stars 
BD+53\,340, GSC\,03671-00025 and GSC\,03671-00791
were used as standards. 

Observations of this symbiotics show periodic wave-like variation -- 
a typical feature of a quiescent phase. The minima profiles and their 
depths are different during different orbital cycles. Recently, 
a secondary minimum at/around the orbital phase 0.5 was seen best 
in the $V$ and $B$ bands. 
As in the case of BF\,Cyg, this behaviour reflects evolution 
in the nebula around the hot ionizing source after the recent active 
phase, which started at 1988.2. Note that also BF\,Cyg underwent a phase of 
activity, which began at 1989.3. 
%
%
%==============================|
%        Fig. 8: AX Per        |
%==============================|
%\vspace*{0.25cm}
\begin{figure}[!ht]
  \centering
  \centerline{\hbox{
  \psfig{figure=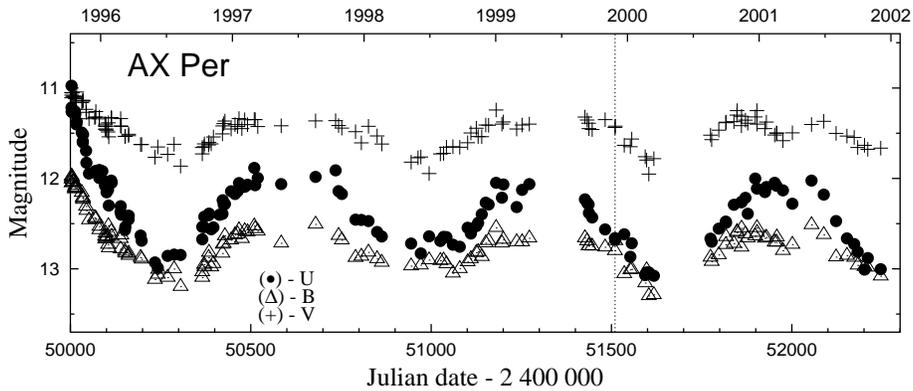,width=12cm,angle=-90}}}
%\vspace{0.25cm}
\caption[ ]{As in Fig. 3, but for AX\,Per.
}
\end{figure}
%
%
%==============================|
%       Table 8: AX Per        |
%==============================|
%
\begin{table}[!ht]
%\small
\scriptsize
\begin{center}
\caption{$U,B,V,R$ observations of AX\,Per}
\begin{tabular}{lccccccc}
\hline
Date & JD~24... & Phase$^{\star}$ & $U$ & $B$ & $V$ & $\Delta R$  & Obs  \\
\hline
 Dec 22, 99 & 51535.406 & 0.859 & 12.618 & 13.055 & 11.634 &  3.438 &SP  \\
 Jan 08, 00 & 51552.377 & 0.884 & 12.867 & 13.012 & 11.650 &  3.452 &SP  \\
 Jan 12, 00 & 51556.247 & 0.890 & 12.717 & 13.006 & 11.565 & --     & SL \\
 Feb 18, 00 & 51593.290 & 0.944 & 13.070 & 13.158 & 11.760 & --     & SL \\
 Feb 22, 00 & 51597.367 & 0.950 & 13.040 & 13.013 & 11.797 &  3.582 &SP  \\
 Feb 28, 00 & 51603.304 & 0.959 & 13.040 & 13.300 & 11.952 & --     & SL \\
 Mar 13, 00 & 51617.313 & 0.980 & 13.077 & 13.288 & 11.781 &  3.547 &SP  \\
 Aug 17, 00 & 51774.452 & 0.211 & 12.666 & 12.874 & 11.522 & --     & SL \\
 Aug 21, 00 & 51777.518 & 0.215 & 12.701 & 12.920 & 11.571 &  3.334 &SP  \\
 Sep 09, 00 & 51797.394 & 0.245 & 12.556 & 12.847 & 11.466 & --     & SL \\
 Sep 29, 00 & 51816.614 & 0.273 & 12.478 & 12.735 & 11.384 &  3.198 &SP  \\
 Sep 30, 00 & 51817.625 & 0.274 & 12.490 & 12.647 & 11.378 &  3.153 &SP  \\
 Oct 21, 00 & 51839.340 & 0.306 & 12.292 & 12.722 & 11.373 & --     & SL \\
 Oct 29, 00 & 51847.464 & 0.318 &  --    &  12.60 & 11.30  & --     & R  \\  
 Oct 30, 00 & 51848.462 & 0.319 &  --    &  12.59 & 11.25  & --     & R  \\
 Nov 11, 00 & 51860.285 & 0.337 & 12.263 & 12.763 & 11.416 & --     & SL \\
 Nov 12, 00 & 51860.629 & 0.338 & 12.26  & 12.6   & 11.31  &  3.13  &SP::\\
 Nov 22, 00 & 51871.386 & 0.353 & 12.215 & 12.617 & 11.356 & --     & SL \\
 Nov 29, 00 & 51877.606 & 0.363 & 12.391 & 12.661 & 11.390 &  3.169 &SP  \\
 Dec 19, 00 & 51898.433 & 0.393 & 12.002 & 12.639 & 11.320 & --     & SL \\
 Dec 24, 00 & 51903.309 & 0.400 & 12.112 & 12.547 & 11.250 &  3.057 &SP  \\
 Dec 24, 00 & 51903.372 & 0.400 &  --    & 12.60  & 11.40  &  --    & R  \\
 Jan 15, 01 & 51925.415 & 0.433 & 12.151 & 12.664 & 11.449 &  3.190 &SP  \\
 Jan 19, 01 & 51929.310 & 0.439 & 12.095 & 12.654 & 11.377 & --     & SL \\
 Feb 11, 01 & 51952.315 & 0.472 & 12.049 & 12.708 & 11.520 & --     & SL \\
 Feb 16, 01 & 51957.326 & 0.480 & 12.081 & 12.707 & 11.498 &  3.242 &SP  \\
 Feb 17, 01 & 51958.319 & 0.481 &   --   & 12.65  & 11.50  & --     & R  \\
 Mar 06, 01 & 51975.330 & 0.506 & 12.133 & 12.800 & 11.583 &  3.218 &SP  \\
 Apr 01, 01 & 52001.292 & 0.544 & 12.280 & 12.732 & 11.496 & --     & SL \\
 May 25, 01 & 52054.523 & 0.623 & 12.024 & 12.514 & 11.404 & --     & SL \\
 Jun 27, 01 & 52088.484 & 0.673 & 12.179 & 12.623 & 11.369 &  3.131 &SP  \\
 Aug 01, 01 & 52122.503 & 0.723 & 12.472 & 12.871 & 11.502 & --     & SL \\
 Aug 31, 01 & 52152.550 & 0.767 & 12.664 & 12.854 & 11.532 &  3.296 &SP  \\
 Sep 20, 01 & 52172.608 & 0.796 & 12.726 & 12.875 & 11.547 &  3.318 &SP  \\
 Sep 30, 01 & 52182.537 & 0.811 & 12.811 & 12.962 & 11.658 &  3.441 &SP  \\
 Oct 18, 01 & 52201.487 & 0.839 & 13.009 & 12.976 & 11.634 &  3.401 &SP  \\
 Oct 27, 01 & 52210.409 & 0.852 & 12.9   & 12.99  & 11.678 &  3.434 &SP::\\
 Dec 02, 01 & 52246.315 & 0.904 & 13.004 & 13.083 & 11.665 & --     &SL  \\ 
\hline
\end{tabular}
\end{center}
$^{\star}$ $Min = JD\,2\,436\,673.3 + 679.9\times E$
           (Skopal 1991)
\normalsize
\end{table}

\subsection{IV\,Vir}

$U,B,V$ measurements of IV\,Vir (BD-21\,3873) are listed in Table 9. 
Observation was carried out at the San Pedro Observatory during June 
2001 (Sect. 2). 
Stars HD\,124991 (BD-21\,3877;  $V$ = 8.07, $B-V$ = 1.05, $U-B$ = 0.73) 
and HD\,125081 (MX\,Vir; $V$ = 7.27, $B-V$ = 0.43, $U-B$ = 0.16), were 
used as standards. This star was also observed visualy by one of 
us (AJ) during a period of 1989.0 to 2001.7. 

Figure 9 shows a phase diagram of our observations of IV\,Vir. It 
displays a double-wave variation throughout the orbital cycle. 
The data were folded according to the ephemeris of the inferior 
spectroscopic conjunction of the cool component in the system. The LC 
profile is similar to that measured in the Str\"omgren $y$ band 
(Smith et al. 1997). A maximum amplitude ($\sim$ depth of the primary 
minimum) is about 0.2\,mag, while the secondary minimum is only 
about 0.15\,mag deep. Currently such type of the periodic light 
variations is attributed to the ellipsoidal shape of the cool 
component due to the tidal distortions by the companion. 
A rival interpretation, based on the extent of the nebula, 
was suggested by Skopal (2001). 
\vspace*{10mm}

%
%==============================|
%        Fig. 9: IV Vir        |
%==============================|
%\vspace*{0.25cm}
\begin{figure}[!ht]
  \centering
  \centerline{\hbox{
  \psfig{figure=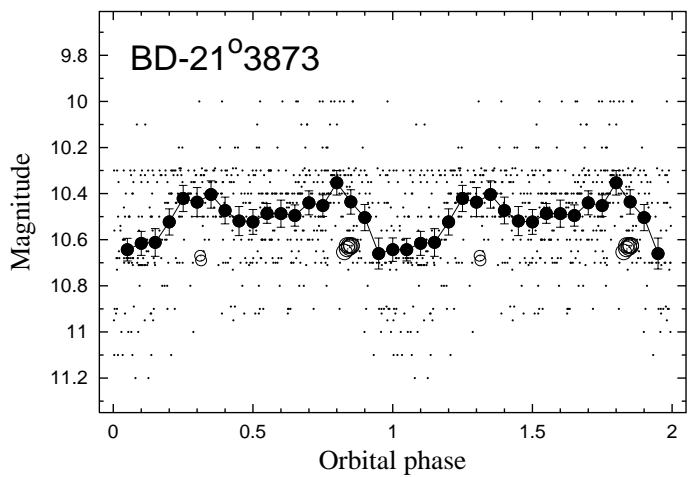,width=9cm,angle=-90}}}
%\vspace{0.25cm}
\caption[ ]{As in Fig. 7, but for IV\,Vir. 
}
\end{figure}
%
%
%==============================|
%       Table 9: IV Vir        |
%==============================|
%
\begin{table}[!ht]
%\small
\scriptsize
\begin{center}
\caption{$U,B,V$ observations of IV\,Vir}
\begin{tabular}{lcccccc}
\hline
Date & JD~24... & Phase$^{\star}$ & $U$ &   $B$   &   $V$  & Obs \\
\hline
 Jun 05, 01 & 52065.681 & 0.827 & 12.525 & 12.040 & 10.653   & M  \\
 Jun 08, 01 & 52068.702 & 0.837 & 12.553 & 12.031 & 10.640   & M  \\
   ------   & 52068.705 & 0.837 & 12.694 & 12.013 & 10.625   & M  \\
 Jun 10, 01 & 52070.733 & 0.845 & 12.716 & 12.059 & 10.634   & M  \\
   ------   & 52070.737 & 0.845 & 12.777 & 12.102 & 10.625   & M  \\
 Jun 12, 01 & 52072.714 & 0.852 & 12.643 & 12.043 & 10.628   & M  \\
   ------   & 52072.718 & 0.852 & 12.579 & 12.021 & 10.624   & M  \\
\hline
\end{tabular}
\end{center}
$^{\star}$ $JD_{\rm sp. conj.} = 2\,449\,016.9 + 281.6\times E$
           (Smith et al. 1997)
\normalsize     

\vspace*{10mm}

\end{table}

\acknowledgements
This research was supported by a grant of the Slovak Academy of Sciences
No. 1157, in part through the Alexander von Humboldt foundation 
(grant SLA/1039115) and by allocation of San Pedro M\'artir observing 
time, from the Czech-Mexican project ME 402/2000 and by the research plan 
J13/98: 113200004 Investigation of the Earth and the Universe.

\end{document}